\documentclass{aa}  

\usepackage{graphicx}
\usepackage{hyperref}
\usepackage{overpic}
\usepackage{upgreek}
\usepackage{txfonts}
\usepackage[version=4]{mhchem}
\newcommand{\ignore}[1]{ }

\begin{document}

   \title{ Tracing bulk elemental ratios in exoplanetary atmospheres with \ce{TiO} chemistry }

    \titlerunning{ Tracing C/O and N/O with TiO }

   \subtitle{}

   \author{Vanesa Ram\'irez\inst{1}\thanks{E-mail: \href{mailto:ramirez@strw.leidenuniv.nl}{ramirez@strw.leidenuniv.nl}}, Alex J. Cridland\inst{1}\thanks{E-mail: \href{mailto:cridland@strw.leidenuniv.nl}{cridland@strw.leidenuniv.nl}}, \and Paul Molli\`ere\inst{2}}

   \institute{$^{1}$ Leiden Observatory, Leiden University, 2300 RA Leiden, the Netherlands\\
   $^{2}$ Max-Planck-Institut f\"ur Astronomie, K\"onigstuhl 17, D-69117 Heidelberg, Germany
              %\email{cvanesa.ramirez@udea.edu.co}
         %\and
          %   Leiden Observatory, Leiden University, 2300 RA Leiden, the Netherlands \\
           %  \email{cridland@strw.leidenuniv.nl}
             }

   \date{}

% \abstract{}{}{}{}{} 
% 5 {} token are mandatory
 
  \abstract{Knowing the bulk elemental abundances of an exoplanetary atmosphere is not an easy task, but it is crucial to understand the formation history of planets. The purpose of this work is to show that the observability of TiO features at optical wavelengths in the transmission spectra of hot Jupiter atmospheres is sensitive to the bulk chemical properties of the atmosphere. For that, we run a grid of chemical models which include TiO formation and destruction, for the ultra-hot Jupiter WASP-19b and an ultra-hot version of HD~209458b. We take into account non-equilibrium chemistry and changes in the temperature and pressure structure of these atmospheres caused by different C/O ratios. We calculate synthetic transmission spectra for these models, and study the relative strengths of TiO and \ce{H2O} features quantitatively. To compare with observations, we use a model independent metric for molecular abundances, $\Delta Z_{\rm TiO-H_2O}/H_{\rm eq}$ that has been previously used in observational studies of exoplanetary atmospheres. We find that with this metric we can differentiate between different chemical models and place constraints on the atmosphere's bulk carbon and oxygen abundance. From chemical considerations we expected that the TiO abundance would depend on the bulk nitrogen, however found that any change to N/H did not result in changes to the resulting TiO. We apply our method to a set of known exoplanets that have been observed in the relevant optical wavelengths and find good agreement between low resolution observations and our model for WASP-121b, marginally good agreement with WASP-79b, WASP-76b, and WASP-19b, and poorer agreement with HD 209458b. Our method can be particularly helpful for indirect studies of the bulk abundances of carbon and oxygen.
}
   %{C/O and N/O are hard to measure}
   %{To show that TiO is senstive to the elemental ratios}
   %{We run a grid of chemical models which include TiO formation and desctruction}
   %{The relative strength of TiO and \ce{H2O} absorption features depend on elemental ratios}


   \maketitle
%
%-------------------------------------------------------------------

\section{Introduction}

The bulk chemical properties of an exoplanetary atmosphere - its carbon-to-oxygen (C/O) and nitrogen-to-oxygen (N/O) ratios - are impacted by the location and timing of the gas accretion phase of its formation \citep[][among many others]{Oberg11,Helling14,Madhu14,Crid16a,Pudritz2018,Cridland2019c,Oberg2019review}. These ratios depend on the chemical structure of the gas and ice throughout the protoplanetary disk \citep{Oberg11,Eistrup2018}, on the chemical properties of the refractory material \citep[i.e. dust and planetesimals,][]{Mordasini16,Cridland2019b} that is accreted directly into the atmosphere, and on the migration history of the growing planet \citep{Madu2014,Crid17}.

Characterising the elemental ratios of exoplanetary atmospheres observationally is crucial to probing the details of planet formation physics. As such, observational programs have already begun using the Hubble Space Telescope \citep[HST, see][]{Madu2011,Kre14,Stevenson2014,Benneke2015,Morley2017,MacDonald2019} and high resolution spectroscopy from the ground \citep{Brogi2013,Birkby2017} to study the chemical properties of hot Jupiters. Furthermore, interferometric studies of directly imaged planets have begun inferring C/O for planets orbiting further from their host star \citep[see for example][]{Gravity2020}. These characterisation studies have long relied on detecting the most abundant carbon and oxygen carrying molecules (ex. \ce{H2O}, \ce{CO2}, \ce{CO}, \ce{CH4}) to determine C/O. Observational programs targeting carbon and oxygen bearing species consist mainly in the near-infrared and are complicated by the presence of clouds \citep[as discussed in][among others]{Benneke2015,Sing15}. Additionally, overlapping spectral features of \ce{CO} and \ce{CO2} along with low spectral resolution \citep{Madu2011b} make determining C/O difficult.

Meanwhile, no observational programs have strongly detected nitrogen-bearing species to determine N/O \citep[but see][for a weak detection of \ce{NH3} and \ce{HCN}]{MacDonald2017}. Given that nitrogen is among the most abundant molecules heavier than helium, its characterization should help to understand the details of planet formation in the same way that carbon and oxygen have. Unfortunately, its primary molecular species - \ce{N2} - is effectively invisible\footnote{It lacks a dipole moment and strong quadrupole moment observable in visible or IR}. Neither of the next two most abundant N-bearing species - \ce{NH3} and HCN - are favoured for hot Jupiter atmospheres with solar C/O. The presence of \ce{NH3} is only expected (in equilibrium chemistry calculations) for low ($< 500$ K) temperatures, while HCN is favoured only for C/O $\geq 1$. In disequilibrium models these molecules can become more abundant due to quenching and photodissociation \citep{Moses2011}, however they never become the primary carrier in hot Jupiter atmospheres.

In this study we posit that less abundant, metal oxides could be used to infer the bulk chemical properties, especially the nitrogen abundance, of hot Jupiter atmospheres. In particular, we focus on titanium oxide (\ce{TiO}) which has strong spectral features in the optical, between 0.4 and 1 microns. This fact has made it a popular target for chemical characterization studies in M- and L-dwarf stars \citep[see for example][]{Veyette2016}. Furthermore, it is known that the abundance of titanium bearing species are strongly linked to the bulk metallicity of M-dwarf stars \citep{Dhital2012,Kesseli2019}, and given the link between bulk metallicity and C/O \citep{Brewer2016b} it should be no surprise that a connection between TiO and bulk chemical properties can be made.

The production and destruction of TiO is discussed in \S \ref{sec:method} and relies on the reaction of elemental titanium with generally less abundant nitrogen and oxygen bearing molecules. We handle its chemistry using the chemical kinetic code VULCAN along with their chemical network that includes the production of TiO. We expect that their abundance, and the subsequent abundance of TiO depend sensitively on C/H, O/H, and N/H, and hence the detection of TiO can be connected, through chemical modelling, to the bulk chemical properties of these abundant elemental species in hot Jupiter atmospheres. We find, however, that changes in N/H do not translate into changes in TiO abundance and hence detections of TiO can not be directly connected to nitrogen abundances - see Appendix \ref{sec:app01}.

Throughout the Method section of this work we will reference N/O as it relates to our prediction that only the relative abundances of those elements are relevant for the chemistry of TiO. Because it turns out that between nitrogen and oxygen, only oxygen is relevant for the TiO abundance we switch to reporting only O/H (relative to solar) in the Results sections as it is more comparable to what can be inferred observationally at this time.

We find that, with the right metric, we can distinguish between different bulk chemical models in the transmission spectra of hot Jupiters. We apply our model to exoplanets with observed transmission spectra covering the necessary wavelength range to deduce whether their inferred C/O and O/H are consistent with chemical modelling. When not included in the observational studies, we assume a solar value for C/O and O/H. We outline our methods in \S \ref{sec:method} and \ref{sec:methspec}, report our results in \S \ref{sec:resratios} and \ref{sec:resspectra}, discuss our results in \S \ref{sec:discussion}, and conclude in \S \ref{sec:conclusion}.

%--------------------------------------------------------------------
\section{Method: Atmosphere chemical model}\label{sec:method}

To model the chemistry occurring in hot Jupiter atmospheres we use the open source chemical kinetic code VULCAN \citep{Tsai2017}. It solves a set of continuity equations for different molecular species, taking into account the production rate, loss rate and transport flux of each species. The set of equations that VULCAN solves are \citep{Tsai2017}:\begin{align}
    \frac{\partial n_i}{\partial t} = \mathcal{P}_i - \mathcal{L}_i - \frac{\partial \phi_i}{\partial z},
\end{align}
where $n_i$ is the number density of the $i$-th species, $\mathcal{P}_i$ and $\mathcal{L}_i$ are the reactions that produce and destroy the $i$-th species. The transport term $\phi_i$ includes both eddy and molecular diffusion, with the former having the form:\begin{align}
    \phi_{i,{\rm eddy}} = -K_{zz} n_{\rm total} \frac{\partial X_i}{\partial z},
\end{align}
where $K_{zz}$ parameterizes the strength of mixing in the vertical ($z$) direction and varies with height. $X_i$ is the abundance of the $i$-th species relative to the total number of particles which is denoted by $n_{\rm total}$. The functional form of the molecular diffusion is taken from \cite{BK1973}.

The production and destruction rates ($k_i$) for the i-th species depend on the local temperature of the gas, and follow the typical modified Arrhenius equation, with the form:\begin{align}
    k_i = A B^T \exp{(-C/T)},
\end{align}
where the parameters $A$, $B$, and $C$ are determined from laboratory experiments, computed numerically, or are estimated. The particular chemical network that we used can be found on the github page of VULCAN.\footnote{\url{https://github.com/exoclime/VULCAN/tree/master/thermo/other_networks}} The network\footnote{\url{TiVNCHO_photo_network_v1025}} includes typical high temperature gas chemistry of carbon, oxygen, and nitrogen-bearing molecules, the oxidizing reactions of vanadium and titanium (discussed more below), and photodissociation reactions of \ce{H2}, \ce{H2O}, \ce{CO}, \ce{CO2}, \ce{CH3}, \ce{CH4}, \ce{C2H2}, \ce{C2H4}, \ce{N2}, \ce{NH3}, and HCN.

\subsection{Elemental abundances}

Our chemical model includes the elements: H, He, C, O, N, V, and Ti, and in general we assume solar abundances for each element (relative to H). In our search through parameter space, we keep N/H constant while varying the carbon and oxygen abundances to achieve our necessary C/O and N/O. As such, our method is sensitive to variations in C/O and O/H. We tested variations in N/H in Appendix \ref{sec:app01} and found no change in the abundance of TiO, discussed below.

The solar abundance of the aforementioned elements are: He/H $= 0.097$, C/H $= 2.78\times 10^{-4}$, O/H $= 4.90\times 10^{-4}$, N/H $= 8.19\times 10^{-5}$ \citep[all from][]{Asplund2009}, V/H $= 9.76\times 10^{-9}$ and Ti/H $= 8.43\times 10^{-8}$ \citep[from][]{Lodders2009}. These elemental abundances result in C/O $= 0.54$ and N/O $= 0.17$, the solar values, and are used as a baseline for our chemical models. We sample C/O $\in [ 0.08, 1.0 ]$ and N/O $\in [ 0.01, 2 ]$ for testing the response of TiO production as a function of bulk chemical composition.

\begin{figure}[t]
\centering 
\includegraphics[width=0.5\textwidth]{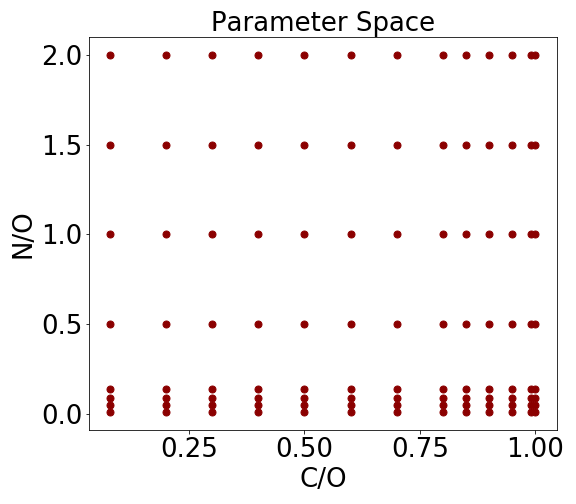}
\caption{Parameter space for our chemical modelling simulation. C/O and N/O ratio span an order of magnitude across values that are attainable from formation models \citep[for example][]{Cridland2019c}.}
\label{fig:parameterspace}
\end{figure}

Our choice of range in C/O was selected to mimic the range of C/O that are attained from formation models of hot Jupiters. \citep{Cridland2019b,Cridland2019c}. In those models, C/O $> 1$ is very difficult to obtain because the gas, ice, and refractory material in protoplanetary disks tend to be more oxygen rich than carbon rich \citep{Mordasini14,Cridland2019b}. Furthermore, one might expect that `carbon-rich' planets (with C/O $>1$) should not have strong spectral features from TiO, as its production should be drastically reduced since most of the oxygen in these types of atmospheres should be in CO.

In the aforementioned hot Jupiter formation models, there are minimal (less than 1\%) differences in the C/H and N/H of the population of synthetic planets. O/H varies by an order of magnitude, which results in N/O variations between 0.01 and 0.1. We extended the range of N/O for our current experiment to account for variations in O/H and N/H that might occur between different stars \citep[i.e. in][]{Brewer2016b}\footnote{In that work O/H varied between 0.6 and 2 $\times$ solar however did not include measurements of N/H}. To fill parameter space we allow C/H to vary more than was seen in the formation models, to similarly account for variations in C/H between different stellar systems\footnote{\cite{Brewer2016b} find C/H variations between 0.3 and 2 $\times$ solar}.

\subsection{Pressure-temperature structure}

We use pre-computed one dimensional pressure-temperature (PT-) curves to model the physical conditions of the atmosphere. These curves were computed using the self-consistent atmospheric petitCODE \citep{Molliere2015,Molliere2017}. In this work we include two models, the first uses the planet properties (mass, radius, surface gravity) of the hot Jupiter HD 209458b, but artificially changes the orbital radius of the planet such that its equilibrium temperature ($T_{\rm eq}$) is 2500~K. This shift in orbital radius changes it classification to an `ultra'-hot Jupiter. We also model the atmospheric structure of the ultra-hot Jupiter WASP-19b (with equilibrium temperature of 2077~K). We use these ultra-hot Jupiter models to guarantee that TiO will not condense into clouds, since our chemical kinetic model does not include cloud formation and heavier elements like titanium should prefer to condense in lower temperature atmospheres. We outline the relevant stellar and planetary properties in Table \ref{tab:planets}.

\begin{table}
    \caption{Relevant stellar and planetary properties}
    \label{tab:planets}
    \begin{tabular}{|c|c|c|}
    \hline 
     & WASP-19b & HD 209458b \\\hline 
    log(g) & 3.16 & 2.97 \\\hline
    T$_{\rm eq}$ (K) & 2077 & 2500 \\\hline
    R$_{\rm planet}$ (R$_{Jup}$) & 1.395 & 1.38 \\\hline 
    M$_{\rm planet}$ (M$_{Jup}$) & 1.114 & 0.69 \\\hline
    Host Type & G8V & G0V \\\hline
    R$_{\rm star}$ (R$_\odot$) & 1.004 & 1.203 \\\hline
    M$_{\rm star}$ (M$_\odot$) & 0.904 & 1.148 \\\hline
    L$_{\rm star}$ (L$_\odot$) & 0.71 & 1.61 \\\hline 
    \end{tabular}
\end{table}

The chemistry used in producing our PT-curves is assumed to be in thermochemical equilibrium, meaning that they are generated by minimizing the Gibbs free energy of the system. This method is very fast, which allows its inclusion into a self-consistent codes like petitCODE. This method, however, differs from the method that we use to predict TiO abundances (described above) - we use the chemical kinetic solver VULCAN to calculate a kinetic steady state for the chemical system. We acknowledge the possible discrepancy between the chemical structure that is used in deriving the PT-curves and the molecular abundances resulting from the chemical kinetic calculation. The base version of VULCAN (ignoring vertical mixing and photodissociation) matches results from the thermochemical equilibrium code TEA \citep{TEAcode,Tsai2017}, but including non-equilibrium effects like vertical mixing and photodissociation should bring the chemical system to a steady state that differs from thermochemical equilibrium. We note that our method focuses on data from transmission spectra which are typically less sensitive to the temperature structure than emission spectra.

%::::::::::::::::::::::::::::::::::::::::::::::::
% FIGURE P-T profile
%::::::::::::::::::::::::::::::::::::::::::::::::

\begin{figure}[t]
\centering 
\includegraphics[width=0.495\textwidth]{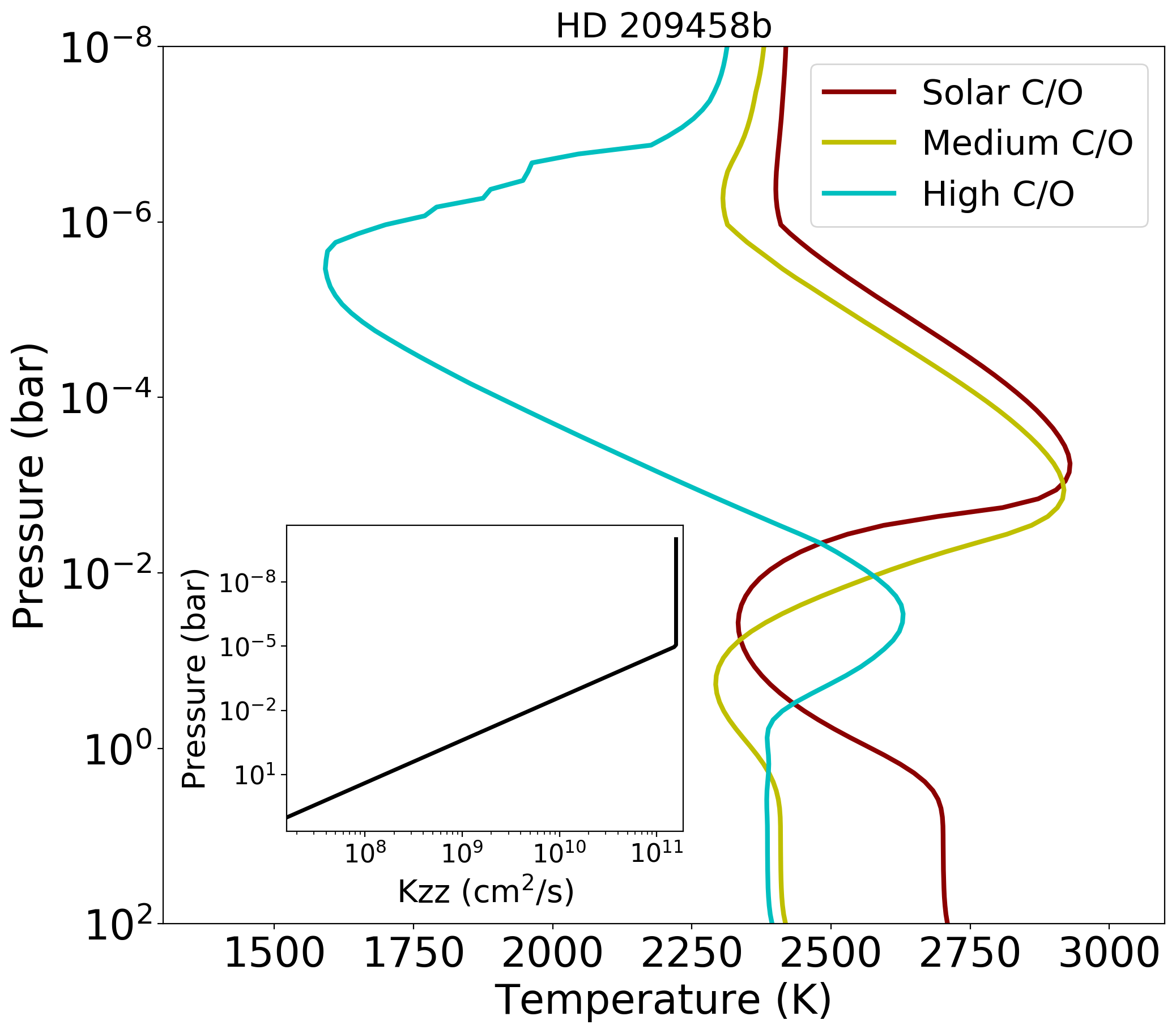}
\caption{Pressure-temperature atmospheric profile for the ultra hot model of HD 209458b. Here we show three models: Solar C/O (=0.54), Medium C/O (=0.8), and High C/O (=1.0). As discussed in \cite{Molliere2015}, thermal inversion occurs in both temperature profiles, even in the high C/O case where TiO is underabundant. In the inset we show the $Kzz$ model used in VULCAN, computed with petitCODE for HD 209458b.}
\label{fig:atm}
\end{figure}

\ignore{ 
\begin{figure}[t]
\centering 
\includegraphics[width=0.49\textwidth]{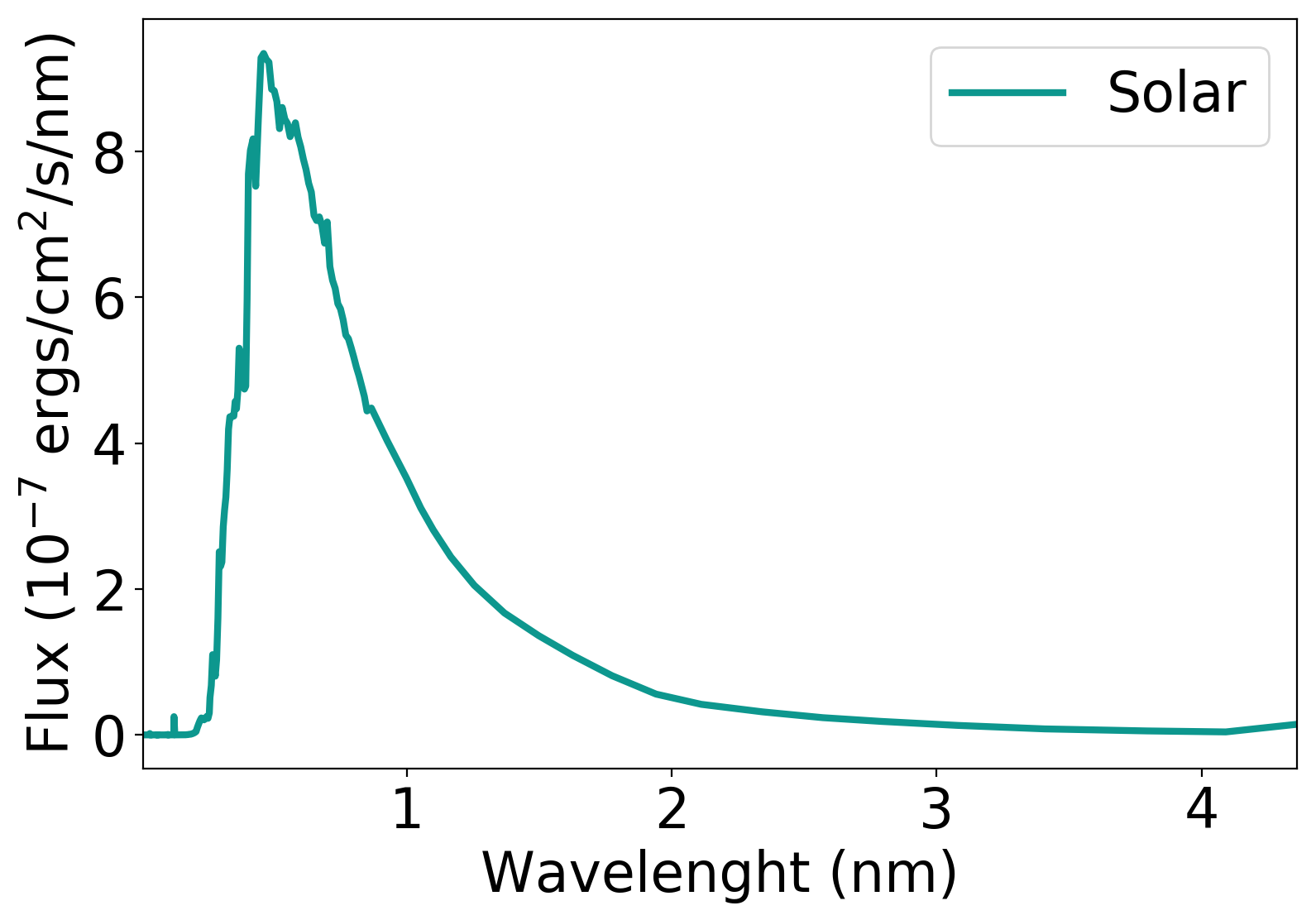}
\caption{Stellar flux used as input for VULCAN to compute our models, corresponding to the solar flux. The two planets we model orbit a solar-type star.}
\label{fig:stellar_flux}
\end{figure}
}

We compute different PT-curves for three different C/O: solar (=0.54), medium (=0.8) and high (=1.0). Different C/O results in a variation in the \ce{H2O} abundance, which can greatly impact the thermal structure in a planetary atmosphere \citep{Molliere2015}. There is a sharp transition in the \ce{H2O} abundance for C/O approaching one - hence atmospheres with C/O $\lesssim 0.7$ should have similar PT-curves \citep{Molliere2015,Gandhi2019}. 

In Figure \ref{fig:atm}, we show the PT-curves for the ultra-hot version of HD 209458b for the three values of C/O. We note that the PT-curves for the solar and medium C/O are very similar, which justifies our use of the same PT-curve over a wide range of C/O when C/O $<$ 0.8. At high C/O the abundance of \ce{H2O} has dropped significantly, and the overall temperature of the atmosphere is reduced\footnote{Note that petitCODE does not include the opacity sources from atomic metal species \citep[for ex. Fe and Ti,][]{Lothringer2020} which could act as important absorbers and heating sources in the upper atmosphere of ultra hot Jupiters.}. In the inset of Figure \ref{fig:atm}, we show the parameterization of the vertical mixing, $K_{zz}$. Often this parameter is kept constant throughout the atmosphere, however we choose to use the pressure-dependent $K_{zz}$ computed by petitCODE. As discussed in the appendix of \cite{Molliere2017}, $K_{zz}$(P) is computed through a combination of mixing length theory \citep[e.g. ][]{KW1990} and fitted results from three dimensional circulation models \citep[i.e. $K_{zz}\propto$ P$^{-1/2}$, see][]{Parmentier2013}.

\subsection{\ce{TiO} production}

Newly added to the VULCAN chemical network are production reactions for TiO. These reactions are largely based on lab experimental data which studied the oxidation of elemental titanium in the gas phase \citep[see][]{Ritter89,Ritter90}. In their work, \cite{Ritter89} found that TiO was easily formed by the oxidation of Ti by \ce{NO}, \ce{O2} and \ce{N2O}; however, TiO was not the end point of their oxidation experiment. They assumed that further oxidation of TiO takes place producing \ce{TiO2}. On top of the TiO reactions in VULCAN, we add two further oxidation reactions to produce \ce{TiO2} based on the kinetic data generated by \cite{Ritter89}. The full set of TiO reactions are: \\
\newline
\noindent
\ce{ Ti + NO <=> TiO + N }\\
\ce{ Ti + O2 <=> TiO + O }\\
\ce{ Ti + N2O <=> TiO + N2 }\\
\ce{ TiO + NO <=> TiO2 + N }\\
\ce{ TiO + O2 <=> TiO2 + O }\\
\newline
\noindent As mentioned above, the oxidizers responsible for the production of TiO are generally underabundant in chemical models of hot Jupiter atmospheres. Hence, we expect that the generation and maintenance of TiO will depend on the bulk chemical composition of the exoplanetary atmospheres.

\begin{figure}
    \centering
    \includegraphics[width=0.5\textwidth]{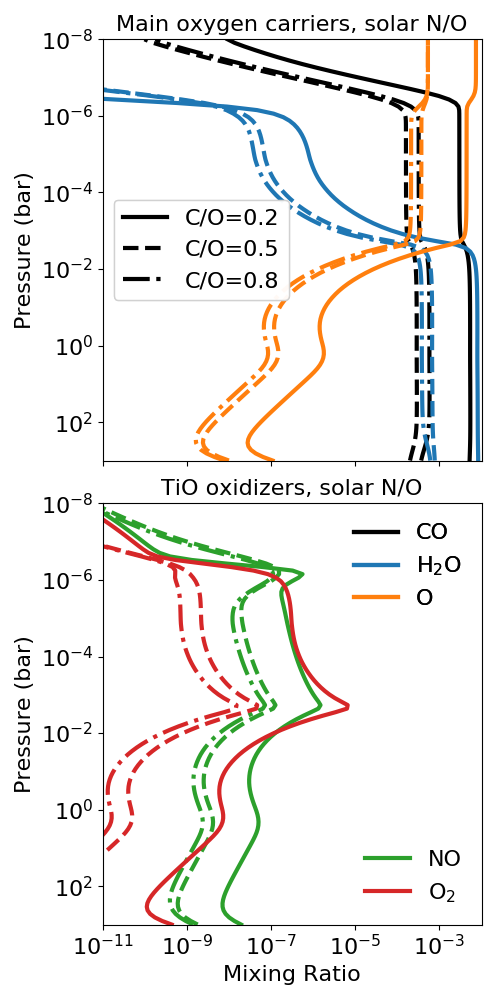}
    \caption{An example of the chemical distribution of the main oxygen carriers in our chemical models. These species are at least two orders of magnitudes more abundant than the oxidizers responsible for the production of TiO. The third TiO oxidizer, \ce{N2O}, is not sufficiently abundant to appear on this figure.}
    \label{fig:example_chem}
\end{figure}

In Figure \ref{fig:example_chem}, we compare the abundances of the main oxygen carriers with the molecules responsible for the generation of TiO. Here we see that these oxidizers are at least two orders of magnitude underabundant compared to the main oxygen carriers. They also vary by between one and two orders of magnitudes over all pressures when C/O is varied by a factor of four. To reiterate, due to these variations we expect the abundance of TiO to vary as we change C/O.

\subsection{Typical chemical structure} \label{typicalchem}

Generally speaking, ultra-hot Jupiters orbit at a typical radius where one might expect photodissociation would dominate the chemical abundances in the upper atmosphere, driven by a high flux of ultra-violet (UV) light. We have assumed a solar-like UV spectrum (based on a PHOENIX model with T$_{\rm eff}$=5700 K) to dictate the flux of UV photons at a given wavelength, and used the photodissocation reactions that are included in the updated version of VULCAN. These reactions include the photodissociation of \ce{H2O}, CO, \ce{CO2}, \ce{H2}, \ce{CH4}, \ce{CH3}, \ce{C2H2}, \ce{C2H4}, \ce{N2}, \ce{NH3} and HCN.

Because of strong photodissociation, the chemistry structure of the atmosphere can typically be thought of being made up of three regions: chemical equilibrium, equilibrium-mixing and radiation-dominated. The first region occurs low in the atmosphere (P $> 1$ bar), where the dissociating photons can not reach the gas and the solution to the chemical model is similar to the thermochemical solution. The second region occurs between $10^{-3}$ and 1 bar, and represents a combination of chemical equilibrium and vertical mixing dictating the abundance of molecular species. In this region we see a fast reduction in \ce{H2O} as it is thermally dissociated into atomic O and H (see Figure \ref{fig:example_chem}), but more stable molecules like CO tend to stay abundant. Very high up in the atmosphere (P $< 10^{-3}$), photodissociation dominates the chemical evolution, and the densities are sufficiently low that recombination is inefficient. In this third region, even the most stable molecules are dissociated, leaving only atomic species in the upper atmosphere.

\begin{figure}
    \centering
    \includegraphics[width=0.5\textwidth]{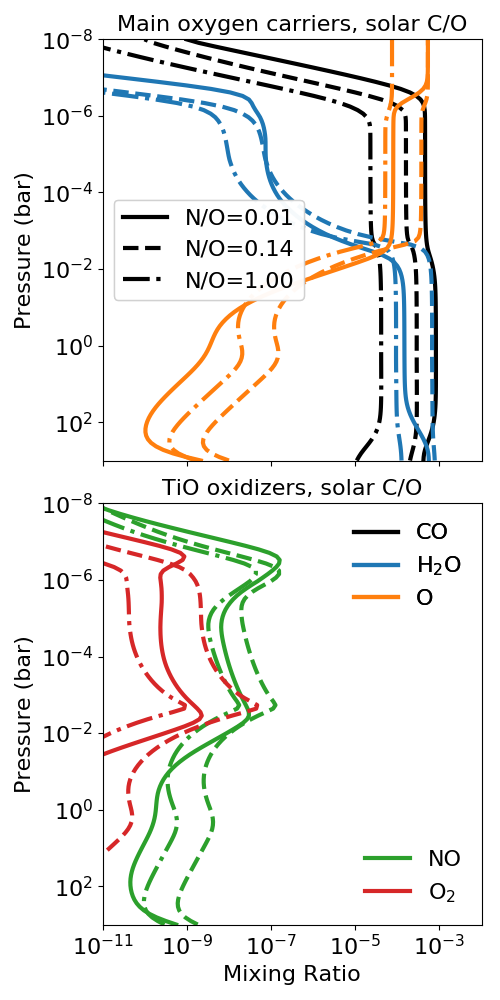}
    \caption{An example of the chemical distribution of the main oxygen carriers in our chemical models as a function of N/O. Similar to Figure \ref{fig:example_chem}, increasing N/O reduces the oxygen abundance available for the production of oxygen carrying species.}
    \label{fig:example_chem2}
\end{figure}

In Figure \ref{fig:example_chem2} we show a similar comparison to Figure \ref{fig:example_chem}, except for variations in N/O with a constant solar C/O. Unlike in Figure \ref{fig:example_chem}, where increases in C/O result in monotonic decreases in TiO oxidizing species, there appears to be an optimal ratio of nitrogen to oxygen ($=$0.14) that results in the highest TiO oxidizer abundances. In our chemical model, we varied N/O by adjusting O/H while keeping N/H constant. We checked what effect changing N/O through variations in N/H had on the resulting abundance of TiO, this test is presented in Appendix \ref{sec:app01}. We find that increases in N/H is not equivalent to a similar decrease in O/H and hence in what follows, variations in N/O should be interpreted primarily as variations in O/H.

\begin{figure}
    \centering
    \includegraphics[width=0.5\textwidth]{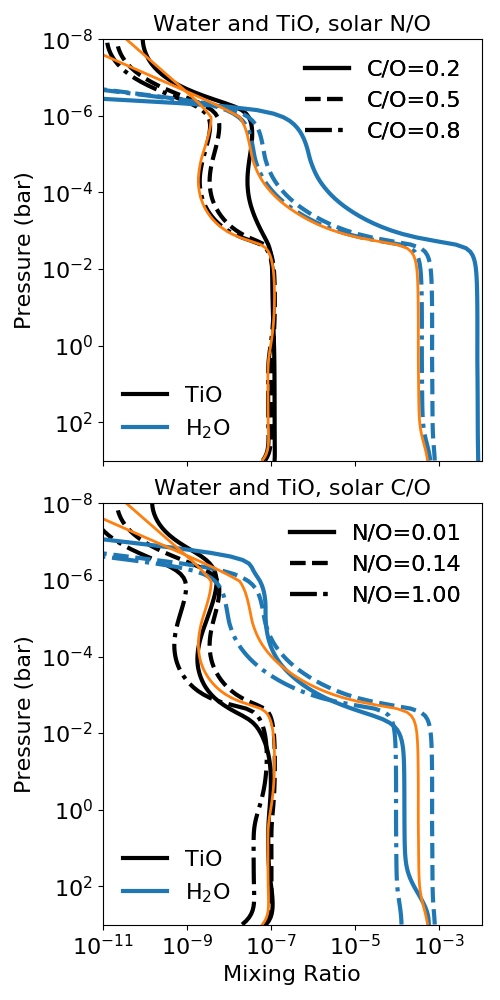}
    \caption{Variation of TiO and \ce{H2O} abundance as a function of both the C/O (top) and N/O (bottom). The abundance of TiO is most sensitive in the region of the atmosphere which is determined by a combination of vertical mixing and photodissociation. For comparison we show the abundances for solar C/O and N/O assuming thermochemical equilibrium with orange lines - these should be compared to the dashed lines in both panels.}
    \label{fig:TiOabun}
\end{figure}

We show in Figure \ref{fig:TiOabun} how these small changes in the abundance of TiO oxidizers impact the steady state abundance of TiO in the atmospheres of the ultra-hot Jupiters. Generally, the largest changes are found in the region of the atmosphere that depends on the combination of vertical mixing and chemical equilibrium. We compare our results to the abundances of TiO and \ce{H2O} derived by thermochemical equilibrium calculations (in orange). We see only small variation between those models and the solar elemental ratio models shown in Figure \ref{fig:TiOabun} (dashed lines). The largest deviations are found in the very upper atmosphere where photodissociation destroys the water that should otherwise be found there. Below, we investigate whether these differences in abundance caused by the bulk chemical properties can be detected by studying the transmission spectra of hot Jupiters. 

\section{Method: Synthetic spectra}\label{sec:methspec}

To compare to observations, we generate synthetic spectra for each of our modeled planets using our chemical models as inputs for petitRADTRANS \citep{Molliere2019}\footnote{https://gitlab.com/mauricemolli/petitRADTRANS}. This radiative transfer scheme assumes that the 1D profile shown in the last section dictates the chemical distribution over all observable latitudes. The code then draws lines of sight through the atmosphere and computes the wavelength-dependent optical depth along these lines of sight. With these optical depths, the code produces a transmission spectrum which describes the apparent size of the planet as it transits in front of its host star.

\begin{figure}[t]
\centering 
\includegraphics[width=0.5\textwidth]{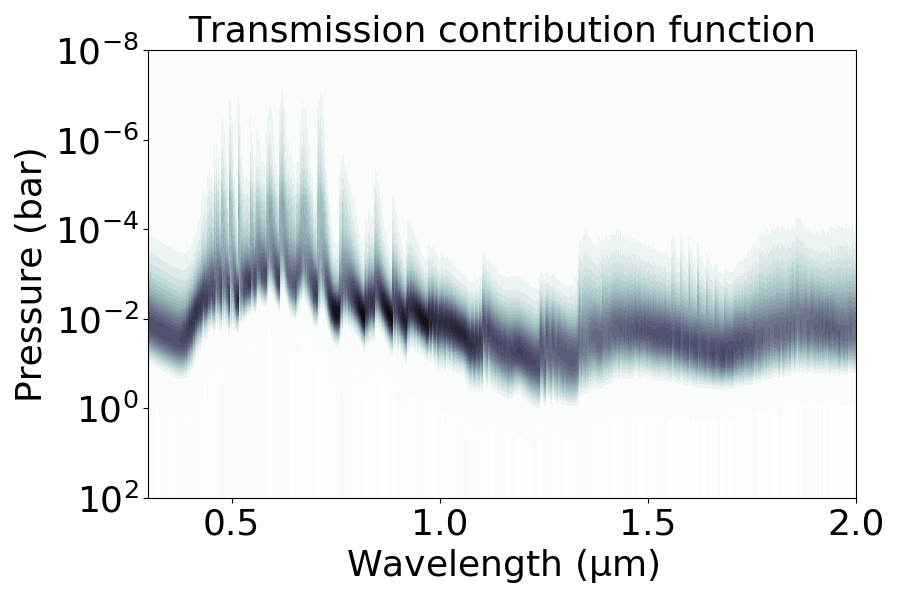}
\caption{Contribution function for the synthetic spectra of the ultra hot version of HD 209458b, in the case of high C/O ratio. The TiO feature primarily comes from the upper levels in the atmosphere, in the region of the atmosphere that depends on both photodissociation and vertical mixing. On the other hand, the H$_2$O feature originates from deeper in the atmosphere, overlapping the chemical equilibrium region.}
\label{fig:contribution_func}
\end{figure}

As the light ray passes through the planetary atmosphere, it passes through different levels of pressures, densities, and as such abundances. In Figure \ref{fig:contribution_func}, we show the contribution function for a typical chemical model of HD 209458b. This function describes from what region in the atmosphere the transmission signal originates. Generally speaking, the TiO signature (between 0.4 and 0.8 $\mu$m) originates from higher in the atmosphere than the 1.4 $\mu$m \ce{H2O} feature.

The species that we consider to generate the synthetic spectra are: \ce{H2}, \ce{H2O}, OH, CO, \ce{CO2}, \ce{CH4}, \ce{C2H2}, HCN, \ce{NH3}, TiO and VO. We also consider Rayleigh scattering for \ce{H2} and He, as well as collision induced absorption for the \ce{H2}-\ce{H2} and \ce{H2}-He pairs. The molecular abundances computed by VULCAN in one-dimension were used as inputs for petitRADTRANS.

%\subsection{Input Parameters}
%% generally speaking editors don't like tiny subsections like this - perfectly fine to have this paragraph finished the method section though
For calculating synthetic spectra, we need to define the planetary radius, planet’s gravity and pressure (at the defined planetary radius), the pressure-temperature profile, the mean molecular weight and the chemical abundance, in units of mass fractions (the former two calculated from VULCAN).

\section{Results: Exploring C/O and O/H parameter space}\label{sec:resratios}

\subsection{Dependence of TiO abundance}

\begin{figure}[t]
\centering 
\includegraphics[width=0.5\textwidth]{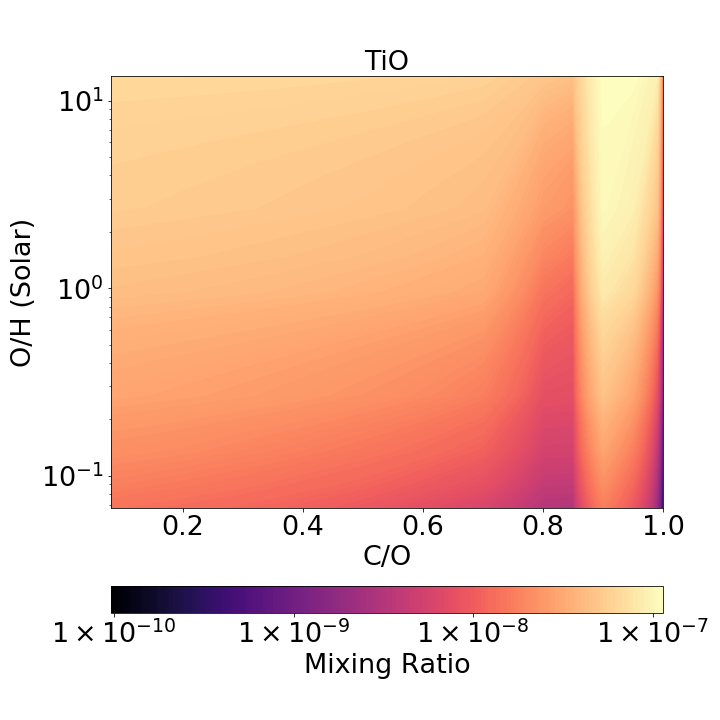}
\caption{Abundance of TiO as a function of C/O and O/H for the ultra-hot model of HD 209458b, averaged through a pressure range of $10^{-4}$ and $10^{-2}$ bar}. There is a smooth decrease in TiO abundance for increasing C/O and decreasing O/H until we switch our PT-curve to the `high C/O' model.
\label{fig:colormapTiO}
\end{figure}

Before directly comparing our synthetic spectra to observations, we first explore the impact of the bulk chemical composition to the abundance of TiO. In Figure \ref{fig:colormapTiO}, we show a colour map for the abundance of TiO varying both C/O and O/H ratios. The abundance that we show is an average across the pressure range of 10$^{-4}$ to 10$^{-2}$~bar, which we assume describes the abundance that will most impact the observed TiO features from transmission spectra (recall Figure \ref{fig:contribution_func}).

We find a smooth transition in the abundance of TiO for C/O $= 0.8$, with the abundance reducing for increasing C/O and decreasing O/H. Above C/O $= 0.85$, we change our PT-curve to the `High C/O' structure shown in Figure \ref{fig:atm}, which results in the discontinuity of TiO abundance that occurs for models with C/O $=0.85$. For models with C/O $> 0.85$, the abundance of TiO again reduces for increasing C/O with the abundance dropping significantly at C/O $= 1$. 

%::::::::::::::::::::::::::::::::::::::::::::::::
% FIGURE TiO colormap
%::::::::::::::::::::::::::::::::::::::::::::::::

Our result of higher TiO abundances for C/O $>0.85$ was unexpected, as equilibrium chemical calculations with high C/O generally result in low TiO abundance. This discrepancy can be understood in context of the chemical models of \cite{Lothringer2018}, who compute TiO abundances over a wide range of pressures and temperatures. For pressures between $10^{-4}$ and $10^{-2}$ bar, our low and medium C/O atmospheric models has gas temperatures between 2500 and 2800 K. At these temperatures, \cite{Lothringer2018} found that equilibrium abundances of TiO were depleted by an order of magnitude because of thermal dissociation. In our high C/O model, the temperatures in the mentioned pressure range is between 1800 and 2400 K, since there is no high thermal inversion in that model. As a result, the chemical system is sufficiently warm that TiO does not condense into clouds, but cold enough that TiO is not thermally dissociated. This stability against thermal dissociation in the higher atmosphere preserves TiO even in chemical models with C/O $=1$, where equilibrium models would otherwise predict significantly lower abundances.

\subsection{Implication for transmission spectra}

%::::::::::::::::::::::::::::::::::::::::::::::::
% FIGURE Transmission spectra
%::::::::::::::::::::::::::::::::::::::::::::::::
\begin{figure}[t]
\centering 
\begin{overpic}[width=0.5\textwidth]{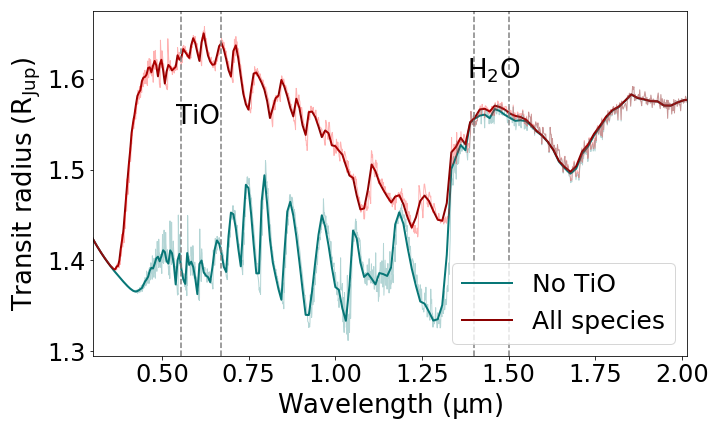}
%\put(25,35){\large TiO}
%\put(67,50){\large \ce{H2O}}
\end{overpic}
\caption{Transmission spectrum for the ultra hot model of HD 209458b. We considering all the species in our chemical model (red line) and compare to a sample spectrum derived for this model but with TiO removed (blue line). The vertical dashed lines denote particular wavelength ranges that we will use in our analysis to compare to observations.}
\label{fig:spectrum}
\end{figure}

TiO has a large number of electronic transitions in the optical (0.4-0.8 $\mu$m) that can easily be detected in exoplanetary atmospheres. In Figure \ref{fig:spectrum}, we compare a transmission spectrum for our model HD 209458b atmosphere with and without TiO. There is a very clear difference between these spectra when we remove TiO. Additionally in the figure, we show the wavelength range that we sample to produce an average transmission signal strength, used in our following analysis (vertical dashed lines).

To compare our chemical models to observations, we average the signal between the vertical dashed lines shown in Figure \ref{fig:spectrum}. We then compare the difference in the signal strength due to TiO ($Z_{\rm TiO}$) with the signal strength caused by \ce{H2O} ($Z_{\rm H_2O}$). This difference is normalized by the equilibrium scale height of the atmosphere ($H_{\rm eq} = kT_{\rm eq}/\mu g$, where $\mu$ is the mean molecular weight of atmospheric particles, assumed to be 2.3 times the mass of the proton, and $g$ the gravity), to act as a metric for molecular abundances in the atmosphere \citep[see for example][]{Sing15}.

%%% this is a very nice addition!
The reason for deriving the metric in this way is as follows: the effective altitude of the atmosphere that is probed by a wavelength $\lambda$ in transmission is estimated by \citep{Lecavelier2008}: \begin{align}
    Z(\lambda) = H_{\rm eq}~ \mathrm{ln}\left(\upxi_{abs}P_{z=0}\sigma_{abs}(\lambda)/\uptau_{\rm eq}\times\sqrt{2\pi R_{\rm p}/kT\mu g}\right)
\end{align}
With $\sigma_{\rm abs}$ and $\upxi_{\rm abs}$ the cross section and abundance of the dominant absorbing species, $P_{z=0}$ is the pressure at the reference altitude, $\uptau_{eq}$ is the optical thickness at the effective radius, and $R_p$ is the planetary radius. The atmospheric scale height is derived for the equilibrium temperature for the planet and the surface gravity. In this way, normalizing by $H_{\rm eq}$ removes from the metric a dependence on the mass and radius of the planet, allowing for a wide range of planets to compare. For two different species, (i.e. TiO and \ce{H2O}) we then have:\begin{align}
%\frac{ Z_{\rm TiO} - Z_{\rm \ce{H2O}}}{H_{\rm eq}} = {\rm ln}(\upxi_{\rm abs,TiO}\sigma_{\rm abs,TiO}) - {\rm ln}(\upxi_{\rm abs,\ce{H2O}}\sigma_{\rm abs,\ce{H2O}})
\frac{\Delta Z_{\rm TiO-H_2O}}{H_{\rm eq}} \equiv \frac{ Z_{\rm TiO} - Z_{\rm \ce{H2O}}}{H_{\rm eq}} = {\rm ln}\left[\frac{\upxi_{\rm abs,TiO}\sigma_{\rm abs,TiO}}{\upxi_{\rm abs,\ce{H2O}}\sigma_{\rm abs,\ce{H2O}}}\right]
\label{eq:metric}
\end{align}
We note that this metric depends on both the relative abundance of molecules and their cross section at the selected wavelengths. The latter only weakly changes between models since different lines-of-sight probe different pressure and temperature ranges resulting in slight changes to cross sections. Comparing this metric between models and observations is most strongly dependent on the relative abundances along a line of sight. Such a comparison can be made between different planets and chemical models since the terms related to the physical properties of the planet are removed.

\begin{figure*}
\centering 
\includegraphics[width=\textwidth]{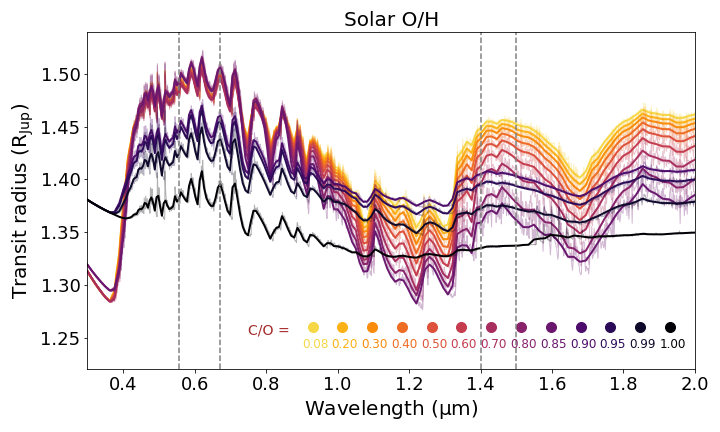}
\caption{Transmission spectra for the ultra hot model of HD 209458b as a function of C/O. The spectra are first computed at high resolution, and are then convolved to a resolution $\lambda/\Delta \lambda=100$. We leave the high resolution spectra as faded curves in the background.}
\label{fig:spectraCO}
\end{figure*}

In Figure \ref{fig:spectraCO}, we show different synthetic spectra that result from different chemical models with varying C/O ratio, for a constant solar O/H. The different C/O are denoted by the changing colour scale, with darker colours representing increasing C/O. While only small changes in the strength of the TiO feature occur up to C/O $=0.8$, the water features becomes increasingly weaker. Once the physical structure of the atmosphere is changed, however, the water feature becomes a stronger transmission feature while the TiO feature begins to be suppressed. The jump in features is related to the change in the physical structure, since the colder upper atmosphere results in a lower scale height. The lower scale height and temperature produces higher densities higher in the atmosphere, allowing for an apparently larger water transmission signal.  For models with high C/O, the specific structure of the water feature does tend to flatten out since the abundance of the molecule is reduced in these models.

\begin{figure*}
\centering 
\includegraphics[width=\textwidth]{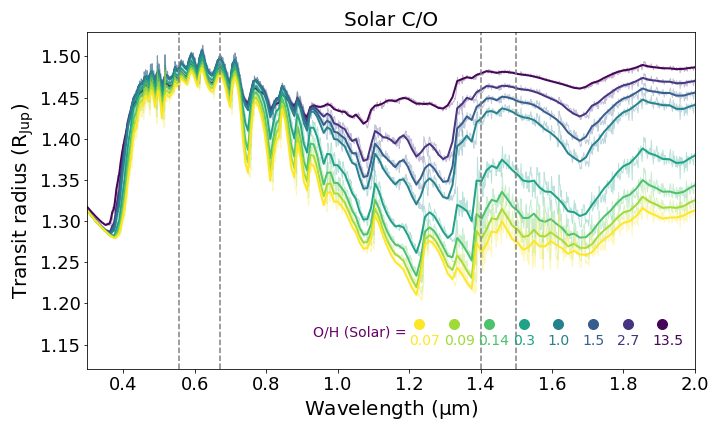}
\caption{Same as Figure~\ref{fig:spectraCO}, but as a function of O/H.}
\label{fig:spectraNO}
\end{figure*}

Similarly, in Figure \ref{fig:spectraNO} we show the variation in the transmission spectra due to changes in O/H, for solar C/O. In this case, the strength of the TiO feature is nearly independent with O/H, which comes from the fact that the abundance of TiO does not vary strongly with O/H. However, since the \ce{H2O} feature is strongly dependent on O/H, the difference in the height of the transmission spectral features will still act as a good metric for determining the bulk properties. While we test O/H depletion by up to a factor of two (N/O enhancement of a factor of two), it is unclear (both observationally and theoretically) whether these elemental ratios could indeed be this large. 

\ignore{ 
For Jupiter, its recent observation of oxygen abundance \citep[O/H $\sim$ 2.7 $\times$ solar; see][]{Li2020} along with its nitrogen abundance \citep[N/H $\sim$ 4 $\times$ solar; see][]{Asplund2009} allows us to estimate N/O $\sim$ 0.25. Given its abundance in protoplanetary disks, it would be difficult to imagine a scenario where oxygen is less abundant than nitrogen. Regardless it is worth testing large N/O since these elemental ratios have not been constrained by observations.
}

In both figures, we can see that the relative heights of the TiO and \ce{H2O} features vary as we change the bulk chemical properties of the atmosphere. We choose to quantify our analysis by using the metric mentioned above, as it should encode the necessary information that would be available to low resolution chemical characterization studies of transiting hot Jupiters.

%::::::::::::::::::::::::::::::::::::::::::::::::
% FIGURE TiO vs C/O 
%::::::::::::::::::::::::::::::::::::::::::::::::

%::::::::::::::::::::::::::::::::::::::::::::::::
% FIGURE TiO vs N/O 
%::::::::::::::::::::::::::::::::::::::::::::::::

\section{Results: Comparing models to observations}\label{sec:resspectra}

%::::::::::::::::::::::::::::::::::::::::::::::::
% FIGURE TiO vs C/O & N/O
%::::::::::::::::::::::::::::::::::::::::::::::::

\begin{figure*}[h!]
\centering 
\includegraphics[width=\textwidth]{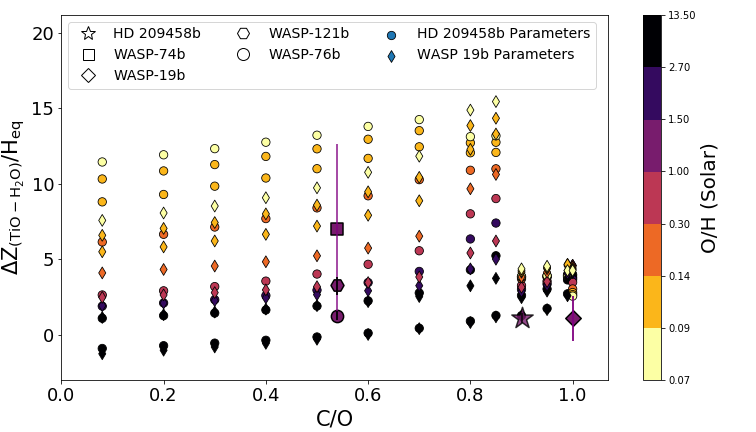}
\caption{The abundance metric for TiO as a function of C/O and O/H ratios. For C/O $\geq$ 0.9 the lower temperatures at lower pressures compress the upper atmosphere. This compression, when normalized by H$_{\rm eq}$ (which is unaffected by C/O) results in the rapid shrink in the metric. The circles and thin diamonds are our physical and chemical models which are compared with observations of hot Jupiters HD 209458b \citep[star; ][]{HD209}, WASP-74b \citep[square; ][]{WASP74b}, WASP-19b \citep[diamond; ][]{WASP19b2013,WASP19b}, WASP-121b \citep[hexagon; ][]{WASP121b}, and WASP-76b \citep[octagon; ][]{WASP76b}.}
\label{fig:comparison}
\end{figure*}

In Figure~\ref{fig:comparison}, we plot $\Delta Z_{\rm TiO - H_2O}/H_{\rm eq}$ as a function of C/O and O/H ratios for our HD 209458b model, as well as the set of models computed with the planetary parameters for WASP-19b. The colour of the points denote the models' O/H, while their position on the x-axis denotes their C/O. We note that in general these points are well separated in the solar and medium C/O models, while these points are less separated in the high C/O model. Similarly, the results from separate atmospheric models (HD 209458b and WAPS-19b) give nearly the same value for the metric in equation \ref{eq:metric} - confirming that this metric should remain consistent while studying a wide range of planetary atmospheres.

As we saw with the raw spectra, the difference in the height of the TiO and \ce{H2O} features increases as C/O increases for lower values of C/O. This trend changes abruptly when we change to the  high C/O model, as the resulting PT-curve has lower temperatures at low pressures. This abrupt change is expected from chemical and physical models of exoplanetary atmospheres \citep[as seen in][]{Molliere2015,Gandhi2019}. With this high C/O model, the specific features of both TiO and \ce{H2O} are suppressed and the difference between their signal strength becomes effectively degenerated with O/H. 

In the solar and medium C/O atmospheric models, decreasing O/H will increase the difference in strength between the transmission features of TiO and \ce{H2O}. As we saw in Figure \ref{fig:spectraNO}, this change in signal comes from a reduction in the strength of the \ce{H2O} signal as O/H decreases rather than a change in the TiO abundance. Regardless, the metric from equation \ref{eq:metric} is a strong tracer of the TiO / \ce{H2O} abundance ratio, which will help to constrain the oxygen abundance in exoplanetary atmospheres.

We compare our models with optical / near-infrared observations of ultra-hot / hot Jupiter atmospheres. For some of these planets, there are direct signals of TiO which are used to derive the abundance of molecules in their atmospheres, as is the case for WASP-19b \citep{WASP19b2013,WASP19b}, WASP-74b \citep{WASP74b}, and WASP-121b \citep{WASP121b}. While in HD 209458b, \cite{HD209} inferred an upper limit for the TiO abundance after removing the transmission signal coming from strong sodium and potassium lines. Below, we outline comparisons between the inferred chemical properties of these planets and our chemical models. When the oxygen abundance is not presented we assume a solar O/H.

\subsection{WASP-121b}

\cite{WASP121b} studied the atmosphere of WASP-121b with the Hubble Space Telescope, and directly detected TiO and \ce{H2O} absorption. They report an abundance ratio of TiO/\ce{H2O} $\sim 7\times$ solar relative to results of thermochemical equilibrium calculations, but do not report specific C/O and O/H. If we assume that both C/O and O/H are solar, then their reported transmission signal is consistent with our synthetic spectra results. This consistency suggests that the non-detection of TiO (replaced with VO) reported by \cite{WASP121b} can be explained by non-equilibrium processes like vertical mixing or photodissociation - which we include in our chemical model - rather than a variation in the bulk chemical properties away from solar.

We note that the work of \cite{WASP121b} is an excellent observational experiment for our purposes, as it provided a wide wavelength range which included wavelengths relevant for both TiO and \ce{H2O} absorption. More observations of this kind would greatly improve our comparisons to observations. Interestingly, further observations of this ultra-hot Jupiter have brought into question the existence of TiO in its atmosphere. Both \cite{WASP121b2018} and \cite{WASP121b2020} report non-detections for TiO in new HST transmission spectra and high resolution cross-correlation studies, respectively. 

If we use the reported data of \cite{WASP121b2018}, the metric $\Delta Z_{\rm TiO - H_2O}/H_{\rm eq} \sim -0.17$ when we average between wavelengths 0.555 $\mu$m - 0.6721 $\mu$m for $Z_{\rm TiO}$. This result is consistent with the aforementioned works, which conclude that TiO is not detected in the atmosphere of this planet. This measurement could be consistent with a solar C/O and super-solar O/H ($\sim$10 $\times$ solar); however, given that \cite{WASP121b2020} report non-detections of both TiO and VO, our method is questionable for such weak signals.

Even more recently, \cite{WASP121b2020b} have completed a molecule and atomic survey of WASP-121b using high resolution spectroscopy. They similarly find non-detections of TiO as well as Ti suggesting that a cold trap could be responsible for a global depletion in titanium in the atmosphere. They predict that \ce{TiO2} could represent an important absorber in the atmosphere, but without an accurate line-list its detection with high resolution spectroscopy will be difficult. We include \ce{TiO2} in our chemical model and find that its abundance is - at most - roughly an order of magnitude less abundant than TiO, and strongly depends on the bulk O/H. It remains to be see what role \ce{TiO2} will play in probing the chemical structure of exoplanetary atmospheres.

\subsection{WASP-74b}

\cite{WASP74b} studied the transmission spectra of WASP-74b from between 500 nm and 2150 nm. They found that their best fit for their optical data was a solar C/O, with the inclusion of TiO and VO. However, a higher C/O model ($=1.1$) gives a better fit to their data at $\lambda \sim 1625$ nm (see their Figure 8). A higher C/O would make their measurement of $\Delta Z_{\rm TiO - H_2O}/H_{\rm eq}$ (square point in Figure \ref{fig:comparison})  more consistent with our results, since a horizontal shift to C/O $\sim 0.8$ would make their estimation agree with our models - as we assumed a solar O/H. 

The wavelength range around 1600 nm has also been suggested as a region to search for nitrogen-bearing species like \ce{NH3} and \ce{HCN} by \cite{MacDonald2017}. As such, a possible explanation for the discrepancy seen in our models, with respect to the measurement of \cite{WASP74b}, is a high bulk nitrogen abundance in its atmosphere. If the oxygen abundance is reduced by a factor of a few, then the combination of a super-solar C/O and reduced O/H would make their measurement more consistent with our models.

\subsection{WASP-19b}

\cite{WASP19b} studied the optical transmission spectra of WAPS-19b and suggested that many features were due to the existence of TiO. This conclusion was in contrast with the work of \cite{WASP19b2013}, who detected water absorption in the atmosphere of WASP-19b, therefore did not expect a high C/O ratio, and suggested an absence of TiO. Meanwhile, \cite{Madhusudhan2012} favours a high C/O ($\geq1$) for this planet founded on ground-based and \textit{Spitzer} photometric data (we set C/O $=1$ in Figure \ref{fig:comparison}). We used the TiO detection of \cite{WASP19b}, and the water detection near 1.4 $\mu$m of \cite{WASP19b2013}, to compute the metric for Figure \ref{fig:comparison}.

We find that the measurement of $\Delta Z_{\rm TiO - H_2O}/H_{\rm eq}$ for WASP-19b falls under the results of our chemical model. One way to overcome the discrepancy between our models and the observational work, as well as explain the observations of both \cite{WASP19b2013} and \cite{WASP19b}, would be if the oxygen abundance of the planet is enhanced with respect to the solar value of O/H. In this way, the diamond point on Figure \ref{fig:comparison} would shift to the left and the colour of the point would become darker (since C/O will be lower and O/H will be higher). The increased O/H would likely produce a stronger \ce{H2O} signal, which might explain the discrepancy between model and observations in \cite{WASP19b2013} (their bottom left panel of Figure 19). An oxygen enhancement of 2 would help to improve the comparison between observations and our chemical models.

We note that \cite{WASP19b2019} additionally studied the optical transmission spectra of WASP-19b using the Magellan/IMACS spectrograph. They report a nearly featureless transmission spectrum in their 0.45 $\mu$m - 0.9 $\mu$m window - precisely where we would expect to see features from TiO. \cite{WASP19b2019} hypothesise that the TiO signal that is reported by \cite{WASP19b} was actually due to starspots or faculae. Of their multiple transits reported, only one showed convincing evidence of TiO features. This transit also had the strongest evidence of stellar contamination which cast doubt on the detection. Clearly more observations of this planet are needed to conclusively constrain the presence (or not) of TiO. 

\subsection{HD 209458b}

The TiO abundance was inferred by \cite{HD209}, and we use the C/O and O/H as inferred by \cite{HD209ratio}. \cite{HD209ratio} inferred a high C/O (between 0.9 and 1.0), assuming solar O/H (they suggest N/O$\geq1$ to explain the detection of HCN in the atmosphere). According to our chemical models, these ratios are inconsistent with the observed strength in the TiO transmission signal. Indeed \cite{HD209} struggled to find a signature of TiO, and needed to subtract the signal from sodium and potassium to find an evidence for TiO. 

The weak TiO signal is attributed to an under abundance of TiO, by a factor of between $10^{-4}$ and $10^{-3}$ below the solar abundance of titanium. Such a reduction could be contributed to the condensing of the molecule into clouds, followed by it raining out of the upper atmosphere. Since HD 209458b is cooler than the ultra-hot models we are considering here (recall that we used an ultra-hot version of HD 209458b), then it is reasonable to assume that cloud formation could be greatly impacting our results. In fact, three-dimensional global circulation models show that cold traps on the night side of the planet can deplete TiO from the upper atmosphere if it condenses into sufficently large particle size \citep{Parmentier2013}. An accurate treatment of the chemical kinetics and cloud formation would be a useful avenue to further our study of this planetary atmosphere; however, for now, all we can say is that non-equilibrium (gas phase) chemistry and TiO formation is insignificant to explain the detection of TiO.

\subsection{WASP-76b}

\cite{WASP76b} studied the optical and infrared transmission spectrum of WASP-76b between 0.3 - 1.5 $\mu$m with HST STIS and WFC3, as well as the emission spectrum from between 1.2 - 1.6 $\mu$m, along with Spitzer data. We used their reported transmission spectrum data near the 0.6 $\mu$m feature of TiO and around the 1.4 $\mu$m feature of \ce{H2O} to compute the metric for Figure \ref{fig:comparison}. Their fitting suggested a slightly super-solar C/O (but within error matches solar); their preferred O/H is not well constrain, but is marginally sub-solar. Our chemical model actually suggests a super-solar oxygen abundance to explain the metric in Figure \ref{fig:comparison}.

\cite{vonEssen2020} and \cite{Edwards2020} have similarly studied the atmosphere of WASP-76b with HST. \cite{Edwards2020} studied both transmission and emission spectra from 1.1 - 1.6 $\mu$m, and found evidence for both TiO and \ce{H2O} in their fits - albeit with higher TiO abundance (by a factor of 1000) in their emission spectra than in their transmission spectra. Due to their choice of wavelength range, we did not include their data in our analysis of WASP-76b. \cite{vonEssen2020} studied the transmission spectra between 0.3 - 0.9 $\mu$m and found no evidence for TiO - instead, they found evidence for TiH which has transitions over a similar wavelength range (this molecule is discussed in section \ref{sec:titanium}). Because they did not include data from the 1.4 $\mu$m feature of water, we similarly did not include their data in our analysis.

It is curious that the three aforementioned papers result in such different conclusions regarding the abundance of TiO. Some of these discrepancies likely arises from the wavelength range of study, and the fact that emission and transmission spectra study different pressure regions in the atmosphere. Indeed, the work of \cite{vonEssen2020} suggests a TiO abundance gradient through the atmosphere (as predicted by our chemical model), and the possible presence of TiH found by \cite{vonEssen2020} suggests that more chemical modeling is required (discussed below).

\section{Discussion: Can \ce{TiO} be used to constrain elemental ratios?}\label{sec:discussion}

Our primary question in this work is whether we can use TiO as an alternative tracer for the bulk chemical composition of (ultra) hot Jupiters. Detecting TiO has a number of benefits because of its significant number of electron transitions in optical wavelengths - making it more efficiently observable from the ground. This helps chemical characterization studies because ground based observatories are more accessible, and can offer much higher spectral resolution than space based observatories. This study also requires a detection of \ce{H2O} in NIR, which can be more difficult from the ground, but recently has been demonstrated with high resolution spectroscopy and cross-correlation \citep[see for example][]{AF2019}.

From Figure \ref{fig:comparison}, we can see that a single measurement of $\Delta Z_{\rm TiO - H_2O}/H_{\rm eq}$ can at least constrain a maximal C/O if the metric is relatively high (i.e. higher than 5). Otherwise, we found that the metric increases slightly as a function of C/O (up to 0.85 in our model), and hence a single measurement of the metric would struggle to constrain both C/O and O/H simultaneously. A measurement of a nitrogen or carbon bearing species would greatly improve our constraints of the bulk chemical properties of the atmosphere. One of these species, along with the measurement of the water feature that must be made to construct the metric would give us a constraint on one of the elemental ratios, which when combined with the metric could constrain the other ratio.

As previously mentioned, our model turned out to not be sensitive to N/H, as variations in N/O are done by varying O/H - and increasing N/H did not result in any change in TiO abundance. We did not expect this to be the case as two of the three TiO oxidizers are nitrogen bearing species. We found, however, that increases of N/H typically resulted in increases of \ce{N2} abundance, which did not translate into more abundant oxidizers and subsequently TiO. We stress, however, that our metric is sensitive to C/H and O/H and hence can still help to understand the bulk chemical properties of hot Jupiter atmospheres.

While the metric from equation \ref{eq:metric} is meant to be model independent, our assumption that TiO and \ce{H2O} are the primary contributors to the spectrum near 0.6 and 1.4 $\mu$m respectively comes from our atmospheric models. Since our model turned out to be insensitive to N/H, a coincident measurement of a nitrogen bearing species would help to characterize both C/O and N/O. This could be done using HCN in the case of high C/O or with \ce{NH3} in the opposite case. There is already evidence of these molecules in hot Jupiter atmospheres \citep[i.e. in ][]{MacDonald2017} and they both happen to have features in the mid-IR which could be observable with JWST-MIRI. Nitrogen could prove to be a stronger tracer of planet formation as its chemistry is generally much simpler than carbon and oxygen. Indeed, analyzing Jupiter's nitrogen abundance has already been used to place strong constraints on its formation history \citep{Bosman2019,Oberg2019}.

\ignore{ 
Current characterization studies lack a constrain on N/O in most planetary atmospheres, since the detection of nitrogen-bearing species are very difficult. Hence, a detection of TiO, \ce{H2O}, and a carbon bearing species would help to place the first strong constraints on N/O. Nitrogen could prove to be a stronger tracer of planet formation as its chemistry is generally much simpler than carbon and oxygen. Indeed, analyzing Jupiter's nitrogen abundance has already been used to place strong constraints on its formation history \citep{Bosman2019,Oberg2019}.
}

\subsection{Incomplete titanium chemistry?}\label{sec:titanium}

In this work we have focused primarily on the oxidation chemistry that leads to the generation of TiO, further oxidation to \ce{TiO2} and the reverse reactions that would return titanium to its atomic state. Hot Jupiters are additionally hydrogen rich - recently new evidence has arisen for titanium monohydride (TiH) in the atmosphere of a hot Neptune \citep{MacDonald2019}, as well as in WASP-76b \citep{vonEssen2020} - and it is possible that chemical competition exists between TiH and TiO. There is a known transition in M, L, and T dwarf stars where metal monohydrides appear in the atmospheres of the cooler L and T dwarfs, at the expense of metal oxides, which are more abundant in warmer M dwarf stars \citep{Kirkpatrick1999}.

In an equilibrium chemistry calculation, \cite{Helling2019} showed that TiH tends to be underabundant relative to TiO for WASP-18b; however, as we had previously discussed, non-equilibrium chemistry can be important for the chemistry to TiO. Moving forward, a more complete chemical kinetic calculation of titanium chemistry, including the production of TiH, should be done. Because of the abundance of hydrogen in the atmospheres of giant planets, TiH could efficiently compete with TiO for the available titanium. TiH also has strong absorption bands in the optical, and could be confused in the transmission spectra of hot Jupiter atmospheres.

\begin{figure}
    \centering
    \includegraphics[width=0.5\textwidth]{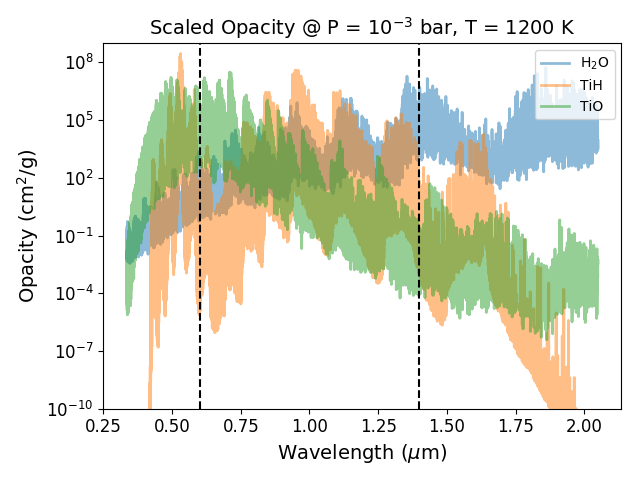}
    \caption{Optical and near-infrared opacities for TiO, TiH, and \ce{H2O}. We scaled the opacity of \ce{H2O} by a factor of $10^{4}$ to account for the larger abundance of \ce{H2O} relative to titanium bearing species in exoplanetary atmospheres. We note the two important wavelength ranges to the metric used in this paper with vertical dashed lines. Data taken from \url{opacity.world} on 12/03/2020.}
    \label{fig:opac}
\end{figure}

In Figure \ref{fig:opac} we illustrate this last point by plotting the opacities of \ce{H2O}, TiO, and TiH. We scaled the opacity of \ce{H2O} by a factor of $10^{4}$ to account for the fact that in atmospheres \ce{H2O} is about four orders of magnitude more abundant than titanium bearing molecules. We see near the wavelength range relevant to our metric that TiH has a feature that is equally strong as TiO, but is much less broad than the feature of TiO. At low spectral resolution and sensitivity it is possible that TiO features could be mistaken for TiH features.

\subsection{Are ultra-hot Jupiters the only place to look for TiO?}

We have intentionally focused on the ultra-hot population of hot Jupiters because of they typically show the best evidence of TiO. The exception of this is HD 209458b, which shows weak evidence of TiO, but is not classified as an ultra-hot Jupiter. It is likely that high ($\geq$ 2000 K) temperatures are needed for strong detections of TiO in the gas phase, since titanium-containing condensates are expected to be formed for temperatures below 2000~K \citep{Lodders2002}. 

While there are currently few planets that are classified ultra-hot Jupiters\footnote{On the NASA exoplanet archive there are currently $\sim 40$ planets with T$_{\rm eq}$ > 2000 K}, the method outlined here can be very useful for characterizing their chemical structure. There is still much to learn about the formation and orbital migration of these ultra-hot Jupiters - and knowing from where they originate will be an important step in understanding their formation history.

Smaller and cooler planets will need other methods of tracing their chemical composition, since we do not expect to find high abundance of TiO in their atmospheres. Metal hydrides may be an interesting avenue of exploration because of their known connection to the lower temperature atmosphere of L and T dwarfs. Surveys of hot Jupiters with high resolution spectroscopy from the ground would be a useful tool for studying atmospheric metal hydrides - nevertheless, we will need a strong modelling effort to interpret their results. Ideally, the exoplanet community can draw from the knowledge already developed for studying dwarf stars\footnote{Which typically rely on equilibrium chemistry}; however, non-equilibrium processes should be considered as hot Jupiters tend to have a bright UV source in their vicinity.

%-----------------------------------------------------------------

\section{Conclusions}\label{sec:conclusion}

In this work we study the connection between detection of TiO and the bulk chemical composition of hot Jupiter atmospheres. We model non-equilibrium chemistry in the atmospheres of two ultra-hot Jupiter models, while varying C/O and O/H using the astrochemical code VULCAN. We account for changes in the temperature and pressure structure of these atmospheres, due to changes in C/O, by considering different structures for low ($<0.8$), medium ($<0.95$) and high ($=1$) C/O. We calculate synthetic spectra for our atmospheric models and compare them with observed transmission spectra that feature TiO detections. 

We use the metric: $\Delta Z_{\rm TiO-H_2O}/H_{\rm eq}$ to quantify our comparisons between model and observations. This metric should be independent with the physical properties of the planet - and only depend on the abundance of a given molecule along the line of sight. With this metric we find that:
\begin{itemize}
    \item WASP-121b is in good agreement with solar values of both C/O and O/H.
    \item WASP-74b is more consistent with a super-solar C/O ($\sim 0.8$) for solar O/H, which partly agrees with the fit done by \cite{WASP74b}. Alternatively, if C/O is solar \citep[as suggested by][]{WASP74b}, then a reduction of O/H by a factor of a few would make the measured strength of the TiO feature consistent with our chemical models.
    \item For WASP-19b, an enhancement in the oxygen abundance by a factor of 2 relative to solar would best explain the measured strength in the TiO feature. This enhancement relative to the reported values of C/O would also help to explain the discrepancy between the studies of \cite{WASP19b2013} and \cite{WASP19b}.
    \item HD 209458b shows only marginal TiO detections which could be caused by TiO condensation and cloud formation. These processes are not included in our chemical analysis, which makes an analysis of HD 209458b difficult in our framework.
    \item WASP-76b would require a higher oxygen abundance than is reported in observational papers to agree with our chemical model. We require super-solar O/H while observational fits prefer sub-solar O/H.
\end{itemize}

We find that TiO detections are a possible tracer of the bulk composition; however, we may be limited to ultra-hot Jupiters with effective temperatures $>$ 2000 K. For complete characterizations, a detection of a nitrogen or carbon carrier is needed to pin down both C/O and O/H. Since our method turned out to not depend on N/H (but rather on O/H by construction) the detection of a nitrogen carrying molecule - like HCN and \ce{NH3} - would be required to give us enough data to constrain the atmospheric N/O, which currently stands as a complete unknown for exoplanetary systems. 

With new missions like the \textit{James Webb Space Telescope} coming online in the near future, exoplanetary science will have a never-before-seen view of exoplanetary atmospheres in the mid-infrared. This will undoubtedly revolutionize our understanding of the bulk chemical composition of exoplanetary atmospheres. However, studies like ours show that efforts in the optical can still give us a unique view of the problem, and further: alternative tracers of bulk chemical composition should be explored. We are confident that the thinking that is deployed in this work will contribute to this task. Furthermore, astrochemical modelling will continue to help plan, search, detect, and characterize more exoplanetary atmospheres as the data becomes available.

\begin{acknowledgements}
     We thank the anonymous referee for their helpful comments that improved the clarity of this manuscript. V.R. is supported by Leiden Observatory through the Oort Scholarship, and was funded as part of the Leiden / ESA Astrophysics Program for Summer students (LEAPS) in the summer of 2019. A.J.C acknowledges support by the European Union ERC grant H2020 ExoplanetBio supervised by Ignas Snellen. P.M. acknowledges support from the European Research Council under the European Union's Horizon 2020 research and innovation program under grant agreement No. 832428. Astrochemistry in Leiden is supported by the European Union A-ERC grant 291141 CHEMPLAN, by the Netherlands Research School for Astronomy (NOVA), and by a Royal Netherlands Academy of Arts and Sciences (KNAW) professor prize.
\end{acknowledgements}

\bibliographystyle{aa} % style aa.bst
\bibliography{mybib.bib} % your references Yourfile.bib
\begin{appendix} %First appendix

\section{Variations in N/O caused by changing N/H}\label{sec:app01}

On the onset of this work, we assumed that N/O would impact the abundance of TiO because of the resulting abundance of the molecules responsible for the oxidation of atomic Ti. Our method involved varying N/O by changing O/H while keeping N/H constant. Here we check whether variations in N/H (for constant O/H) have the same effect on the abundance of TiO.

\begin{figure}
    \centering
    \includegraphics[width=0.495\textwidth]{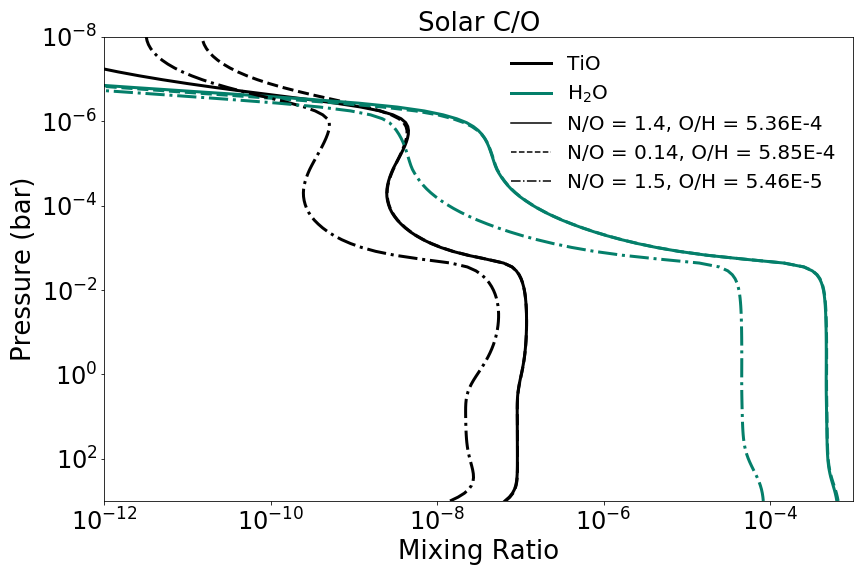}
    \caption{Variation in water and TiO abundances caused by variations in bulk N/O for solar C/O. Here we varied N/O by increasing N/H by an order of magnitude and compared to the models in the main text which kept N/H constant.}
    \label{fig:app01}
\end{figure}

In Figure \ref{fig:app01}, we show the result of increasing N/H by an order of magnitude and comparing its results to two other models used in the main text. We find that increasing N/H by an order of magnitude does not have the equivalent effect on the resulting abundance of TiO as decreasing O/H by an order of magnitude (even though their N/O are the same - see the solid and dot-dashed lines).

The difference in increasing N/H and decreasing O/H can be easily explained by considering the primary nitrogen carrier - \ce{N2}. Increasing N/H by an order of magnitude generally results in an increase in the \ce{N2} abundance, but has a much smaller effect on the TiO oxidizers. An increase in N/H does not translate strongly to an increase in NO abundance (for example) because there is no more available oxygen for its production. However when O/H is decreased, the abundance of NO will also see a decrease in its abundance (and subsequently TiO).

\end{appendix}

\end{document}